\documentclass[12pt]{article}

\usepackage{graphicx}

\def\lsim{\mathrel{\rlap{\lower4pt\hbox{\hskip1pt$\sim$}}
    \raise1pt\hbox{$<$}}}               
\def\gsim{\mathrel{\rlap{\lower4pt\hbox{\hskip1pt$\sim$}}
    \raise1pt\hbox{$>$}}}               

\newcommand{\be}{\begin{eqnarray}}
\newcommand{\ee}{\end{eqnarray}}

\begin{document}

\rightline{\Large Preprint RM3-TH/03-6}

\vspace{1cm}

\begin{center}

\LARGE{New extraction of $\alpha_s(M_Z)$ from proton $DIS$ data\footnote{\bf To appear in Nuclear Physics B.}}

\vspace{1.5cm}

\large{S.~Simula$^*$ and M.~Osipenko$^{**}$}

\vspace{0.5cm}

\normalsize{$^*$Istituto Nazionale di Fisica Nucleare, Sezione di Roma III, I-00146 Roma, Italy\\
$^{**}$Moscow State University, 119992 Moscow, Russia and INFN, Sezione di Genova\\ I-16146 Genova, Italy}

\end{center}

\vspace{1cm}

\begin{abstract}

\noindent An exploratory study for a new determination of the strong coupling constant $\alpha_s(M_Z)$ from existing world data on the proton structure function $F_2^p$ in the $Q^2$-range from $\approx 5$ to $\approx 120 ~ (GeV/c)^2$ is presented. The main features of our approach are: ~ 1) the use of low-order Nachtmann moments evaluated with a direct contribution from data larger than $70\%$ of the total; ~ 2) the inclusion of high-order perturbative effects through the soft gluon resummation technique; ~ 3) a direct control over higher-twist effects; and ~ 4) the independence from any specific choice of the $x$-shape of the input parton distributions. At next-to-leading order we get $\alpha_s(M_Z) = 0.1209 \pm 0.0010 ~ (stat.) \pm 0.0015 ~ (syst.)$ with a significant dependence upon the order of the moment used. Including soft gluon effects we obtain $\alpha_s(M_Z) = 0.1188 \pm 0.0010 ~ (stat.) \pm 0.0014 ~ (syst.)$ with a remarkable better stability against the order of the moment. Our findings are compared with recent $DIS$ determinations of $\alpha_s(M_Z)$ and directions for future improvements are discussed.

\end{abstract}

\vspace{1cm}


\vspace{0.5cm}


\newpage

\pagestyle{plain}

\section{Introduction}
\label{sec:intro}

\indent The coupling constant $\alpha_s$ is the only parameter of Quantum ChromoDynamics ($QCD$) which is our present field theory of the strong interactions. The energy (or scale) dependence of $\alpha_s$ is fixed by the renormalization group equation of $QCD$, while the value of $\alpha_s$ at a given scale must be determined from experiments.

\indent Up to now there are many measurements of $\alpha_s(\mu)$ at different scales $\mu$, coming mainly from analyses of deep inelastic scattering ($DIS$), $e^+ e^-$ annihilation and hadron collider processes as well as from analyses of the heavy quarkonium system. It is well established (cf. Ref.~\cite{Bethke}) that the running of $\alpha_s(\mu)$ is properly predicted by $QCD$ for $\mu \gsim 1 ~ GeV$ up to 4-loop corrections, and therefore it makes sense to extrapolate all the existing measurements of $\alpha_s(\mu)$ to a common reference scale, which is conventionally taken to be the rest mass of the neutral $Z$ boson, $M_Z = 91.1876 ~ (21) ~ GeV$ \cite{PDG}.

\indent The latest world average of $\alpha_s(M_Z)$, based on results of $QCD$ analyses performed at next-to-next-to-leading order ($NNLO$), is $\alpha_s(M_Z) = 0.1183 \pm 0.0027$ \cite{update}. As far as the extraction of $\alpha_s(M_Z)$ from $DIS$ data is concerned, the current average value at next-to-leading order ($NLO$) is $\alpha_s(M_Z) = 0.119 \pm 0.002 ~ (exp.) \pm 0.003 ~ (th.)$ \cite{update}, where the theoretical error includes an estimate of $NNLO$ effects. Recently, two new determinations of $\alpha_s(M_Z)$ have been carried out using the complete $NNLO$ of perturbative $QCD$, obtaining $\alpha_s(M_Z) = 0.1166 \pm 0.0009 ~ (exp.) \pm 0.0010 ~ (th.)$ \cite{SY01} and $\alpha_s(M_Z) = 0.1143 \pm 0.0014 ~ (exp.) \pm 0.0013 ~ (th.)$~\cite{Alekhin}.

\indent The methodology commonly used to extract $\alpha_s(M_Z)$ form $DIS$ data can be summarized into the following steps: ~ 1) fix the perturbative order (and the renormalization scheme) of the analysis; ~ 2) choose a parameterization for the input parton distributions at a given scale $\mu$; ~ 3) fit scaling violations using $\alpha_s(\mu)$ as a free parameter. Finally, one extrapolates $\alpha_s(\mu)$ to $\alpha_s(M_Z)$. Therefore, the main sources of uncertainties are expected to be: ~ i) the effects of high-order perturbative corrections; ~ ii) the dependence upon the specific choice of the $x$-shape of the input parton distributions; and ~ iii) possible effects of higher twists, which can mimic scaling violations in a finite $Q^2$ window. 

\indent The extraction of $\alpha_s(M_Z)$ from scaling violations of the nucleon structure function $F_2^N$ is characterized by a remarkable abundance of data with reasonably small errors. There is however a non-negligible dependence upon the input parton densities, particularly on the input used for the gluon distribution. One can speculate that in the non-singlet channel the so-called log-slope of $F_2^N$ has a reduced dependence upon parton distributions. However, existing data on the non-singlet part of $F_2^N$, which involves the difference between the proton and neutron structure functions, are not very accurate. Alternatively, one can consider the structure function $F_3^{\nu p}$, which is directly a non-singlet quantity; again however the experimental uncertainties on $F_3^{\nu p}$ are typically large (see Ref.~\cite{Kataev} for the latest determination of $\alpha_s(M_Z)$ using neutrino $DIS$ data).

\indent Recently, the strong coupling $\alpha_s(M_Z)$ has been extracted from the scaling violations of truncated moments of the non-singlet part of the structure function $F_2^N$ \cite{Forte}. The truncated moments are determined through a parameterization of $BCDMS$ and $NMC$ data obtained using a neural network \cite{neural}, which allows to retain the full experimental information on errors and correlations. At $NLO$ the authors of Ref.~\cite{Forte} have obtained $\alpha_s(M_Z) = 0.124 ~ _{-0.007}^{+0.004} ~ (exp.) ~ _{-0.004}^{+0.003} ~ (th.)$. The method of Refs.~\cite{Forte,neural} has the nice feature of minimizing many theoretical uncertainties; in particular the neural network parameterization of $F_2^N$ does not rely on any specific choice for the input parton distributions. In order to avoid effects from higher twists, which are contained in the neural network fit, the $Q^2$ range used in Ref.~\cite{Forte} for the extraction of the strong coupling was restricted to $Q^2 > 20 ~ (GeV/c)^2$. In this way however the sensitivity to scaling violations is reduced and this may generate large uncertainties on the extracted value of $\alpha_s(M_Z)$.

\indent The aim of this paper is to present an exploratory study for a new extraction of $\alpha_s(M_Z)$ from proton $DIS$ data. The main features of our procedure are: ~ 1) the use of low-order Nachtmann moments evaluated with a direct contribution from data larger than $70\%$ of the total moment; ~ 2) the inclusion of high-order perturbative effects through the soft gluon resummation technique at next-to-leading-log ($NLL$) accuracy; ~ 3) a direct control over higher-twist effects. As a result, our determination of $\alpha_s(M_Z)$ does not rely on any assumed $x$-shape of the input parton distributions and the $Q^2$-range of our analysis can be extended from $\approx 5$ to $\approx 120 ~ (GeV/c)^2$. At next-to-leading order we get $\alpha_s(M_Z) = 0.1209 \pm 0.0010 ~ (stat.) \pm 0.0015 ~ (syst.)$ with a significant dependence upon the order of the moment used. Including soft gluon effects we obtain $\alpha_s(M_Z) = 0.1188 \pm 0.0010 ~ (stat.) \pm 0.0014 ~ (syst.)$ with a remarkable better stability against the order $n$ of the moment, at least for $n < 8$.

\indent The plan of the paper is as follows. In Section \ref{sec:Mn} low-order Nachtmann moments of the proton structure function are evaluated in the $Q^2$-range $5 \lsim Q^2 ~ (GeV/c)^2 \lsim 120$, following the procedure of Ref.~\cite{JLAB}. In Section \ref{sec:QCD} the precise $Q^2$-range for a leading twist analysis is determined for each moment separately. Then, the $pQCD$ analysis is carried out both at $NLO$ and including soft gluon effects at $NLL$ accuracy. In Section~IV our findings are compared with those of recent $DIS$ determinations of $\alpha_s(M_Z)$ and directions for future improvements are discussed, namely: ~ i) the need for more precise data at least in the $Q^2$ range from $5$ to $15 ~ (GeV/c)^2$, which might come from an upgrade of Jefferson Lab to $12 ~ GeV$ electron beam; ~ ii) the consideration of more Nachtmann moments in our analysis; and ~ iii) the inclusion of all the $NNLO$ effects and of the resummation of large $n$-logarithms at $NNLL$ accuracy. Our conclusions are finally summarized in Section~V.

\section{Evaluation of the Nachtmann Moments}
\label{sec:Mn}

\indent In this Section we review the methodology developed in Ref.~\cite{JLAB} to evaluate the Nachtmann moments of the proton structure function starting from world data. There, all the existing data in $DIS$ kinematics as well as the new results obtained with the $CLAS$ detector at Jefferson Lab in the nucleon resonance production regions, covering a wide $Q^2$-range from $\approx 0.1$ to $\approx 5 ~ (GeV/c)^2$, were considered. For sake of completeness of the present paper we describe here the methodology of Ref.~\cite{JLAB}, but we limit ourselves to the world data set at $Q^2 > 5 ~ (GeV/c)^2$.

\indent As it is well known \cite{Nachtmann}, the Nachtmann moment $M_n(Q^2)$ of the proton structure function $F_2^p(x, Q^2)$ is defined as 
 \be
     M_n(Q^2) \equiv \int_0^1 dx \frac{\xi^{n + 1}}{x^3} ~ F_2^p(x, Q^2) ~ 
     \frac{3 + 3(n + 1) r + n(n + 2) r^2}{(n + 2)(n + 3)}
     \label{eq:Nachtmann}
 \ee
where $n$ is the order of the moment, $\xi = 2 x / (1 + r)$ is the Nachtmann variable and $r \equiv \sqrt{1 + 4 M^2 x^2 / Q^2}$ with $M$ being the nucleon mass. We have chosen the Nachtmann definition of the moments instead of the Cornwall-Norton one, because with the former all the effects arising from target-mass corrections are exactly cancelled out. Hereafter, for sake of simplicity, we will omit the superscript "p" and indicate the proton structure function simply by $F_2(x, Q^2)$. 

\indent Since moments are fully inclusive quantities, Eq.~(\ref{eq:Nachtmann}) receives contributions from inelastic channels as well as from the elastic peak, namely:
 \be
     M_n(Q^2) = M_n^{in}(Q^2) + M_n^{el}(Q^2)
     \label{eq:inel+el}
 \ee
with
 \be
    M_n^{el}(Q^2) = \frac{G_E^2(Q^2) + \tau ~ G_M^2(Q^2)}{1 + \tau} ~ 
    \xi_{el}^{n + 1} ~ \frac{3 + 3(n + 1) r_{el} + n(n + 2) r_{el}^2}{(n 
    + 2)(n + 3)}
    \label{eq:Mn_el}
 \ee
where $G_E(Q^2$ ($G_M(Q^2)$) is the charge (magnetic) Sachs proton form factor,
$\xi_{el} \equiv 2 / (1 + r_{el})$, $r_{el} = \sqrt{1 + 1 / \tau}$ and $\tau \equiv Q^2 / 4 M^2$.

\indent The evaluation of the inelastic moment $M_n^{in}(Q^2)$ involves the computation at fixed $Q^2$ of an integral over the whole range of values of $x$ up to the inelastic pion threshold. For this purpose world data on the structure function $F_2$ from Refs.~\cite{HALLC,BCDMS,E665,NMC,H1,ZEUS,SLAC} and data on the inelastic cross section \cite{Bodek,csworld} are considered in order to reach an adequate kinematical coverage. The integral over $x$ is performed numerically, using the standard trapezoidal method $TRAPER$~\cite{cernlib}. Data from Ref.~\cite{EMC} are not included in the analysis due to their inconsistency with other data sets as explained in detail in Ref.~\cite{MIL91}, and data from Ref.~\cite{WA25,inclusive} are not considered due to the large experimental uncertainties. 

\indent First of all the $Q^2$-range from $5$ to $120$ $(GeV/c)^2$ is divided into bins having a width equal to $\Delta Q^2 / Q^2 = 5 \%$. Bins do not follow one another, but they are located only where data exist. Nevertheless, different bins do not overlap. Then world data are sorted within each $Q^2$ bin and shifted to the central bin value $Q_0^2$, using the fit from Ref.~\cite{Ricco}, which consists of the $SMC$ parameterization of Ref.~\cite{SMC} at $W > 2.5 ~ GeV$ and the $SLAC$ fit from Ref.~\cite{Bodek} for $W < 2.5 ~ GeV$, namely
 \be
    F_2(x, Q_0^2) = \frac{F_2(x, Q^2)}{F_2^{fit}(x, Q^2)} ~ F_2^{fit}(x, 
    Q_0^2) ~ .
    \label{eq:d_nm2}
 \ee
Then the difference between real and bin-centered data
 \be
    \delta^{cent}_{F_2}(x, Q_0^2) = F_2(x, Q^2) \left| 1 - \frac{F^{fit}_2(x, 
     Q_0^2)}{F_2^{fit}(x, Q^2)} \right| ~
 \ee
is added (in quadrature) to the systematic errors of $F_2$.

\indent In order to have a data set dense in $x$, which reduces the error in the numerical integration, we perform an interpolation, at each fixed $Q^2_0$, when two contiguous experimental data points differ by more than $\nabla$.  The value of $\nabla$ is not unique and depends on kinematics. We divide the whole range of $x$ (or $W$) in three intervals: resonance region ($W < 1.8$ GeV), normal $DIS$ region ($W > 1.8 ~ GeV$ and $x < 0.06$) and very low-$x$ region ($x < 0.06$). In the resonance regions, where the structure function exhibits strong variations, we put $\nabla$ equal to $0.01$. Above resonances where $F_2$ is smooth, $\nabla = 0.1$, and finally in the low $x$ region we set $\nabla = 0.005$. Changing these $\nabla$ values by as much as a factor of two produces changes in the moments that are much smaller than the final systematic errors reported in Table~\ref{table:nm}.

\indent To fill the gap between two contiguous data points, $x_a$ and $x_b$, which are distant each other more than $\nabla$, we use the interpolating function $F_2^{int}(x, Q^2)$ defined as the parameterization from Ref.~\cite{Ricco} normalized to the data on both edges of the interpolating range. Assuming that the $x$-shape of $F_2^{int}(x, Q^2)$ is correct, one has
 \be
    F_2^{int}(x, Q^2) = \Pi(Q^2) ~ F_2^{fit}(x, Q^2) ~ ,
    \label{eq:d_nm_i1}
 \ee
where the normalization factor $\Pi(Q^2)$ is defined as the weighted average evaluated using all experimental points located within the interval $\Delta$ around $x_a$ or $x_b$, namely
 \be
    \Pi(Q^2) = \delta_N^2(Q^2) \sum\limits_{i}^{|x_i - x_{a,b}| < \Delta}
    \frac{F_2(x_i, Q^2)}{F_2^{fit}(x_i,Q ^2)} ~ \frac{1}{\left[ 
    \delta_{F_2}^{stat}(x_i, Q^2) \right]^2} ~ ,
    \label{eq:d_nm_i2}
 \ee
where $\delta_{F_2}^{stat}(x_j, Q^2)$ is the statistical error relative to $F_2^{fit}$ and
 \be
    \delta_N(Q^2) = \left\{ \sum\limits_{i}^{|x_i - x_{a,b}| < \Delta} 
    \frac{1}{\left[ \delta_{F_2}^{stat}(x_i, Q^2) \right]^2} \right\}^{-1/2}
    \label{eq:d_nm_i3}
 \ee
is the statistical uncertainty of the normalization. Therefore, the statistical error of the moments calculated according the trapezoidal rule~\cite{cernlib} was increased by adding the linearly correlated contribution from each interpolation interval as follows:
 \be
    \delta^{norm}_n (Q_0^2) = \delta_N(Q_0^2 ) \int\limits_{x_a}^{x_b} 
    dx \frac{\xi^{n + 1}}{x^3} ~ F_2^{fit}(x,Q_0^2) ~ \frac{3 + 3(n + 1) 
    r + n(n + 2) r^2}{(n + 2)(n + 3)} ~ .
    \label{eq:d_nm_i25}
 \ee

\begin{figure}[htb]

\centerline{
\includegraphics[bb = 1cm 6cm 20cm 23cm, scale=0.5]{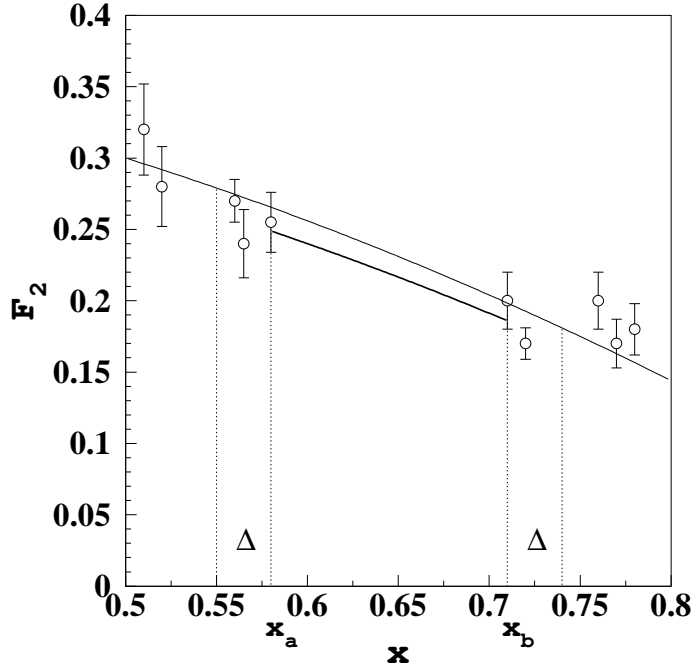}
}

\caption{\label{fig:interp} \small \em Example of the interpolating procedure. The meaning of both the curves and the symbols is described in the text.}

\end{figure}

\indent Since we average the ratio $F_2(x_i, Q^2) / F_2^{fit}(x_i, Q^2)$, the value of $\Delta$ is not affected by the resonance structures in $F_2$, and we fix it by the requirement to have more than two experimental points in most cases, obtaining $\Delta = 0.01$ in the resonance and in the very low-$x$ regions, and $\Delta = 0.1$ in the $DIS$ region. In Fig.~\ref{fig:interp} we show how this interpolation is applied: the thin continuous line represents the original function $F_2^{fit}(x, Q^2)$ and the heavy continuous line represents the result of the interpolation $F_2^{int}(x, Q^2)$. We have also checked that the moments do not show any significant dependence on the specific value of $\Delta$.

\indent To fill the gap between the last experimental point and one of the integration limits ($x_a = 0$ or $x_b = 1$) we perform an extrapolation at each fixed $Q^2$ using $F_2^{fit}(x,Q^2)$ including its systematic error given in Refs.~\cite{Ricco,Bodek,SMC,MIL91}.

\indent The total systematic error consists of genuine uncertainties in the data, as given in Refs.~\cite{HALLC,BCDMS,E665,NMC,H1,ZEUS,SLAC,Bodek,csworld}, and uncertainties in the evaluation procedure. To estimate the first type of error one has to take into account that many data sets have been measured in different laboratories and with different detectors. We assume that different experiments are independent and therefore only systematic errors within one data set are correlated. Thus, an upper limit for the contribution of the systematic error from each data set is evaluated in the following way:

\begin{itemize}

\item{we first apply a simultaneous shift to all experimental points in a data set by an amount equal to their systematic error;}

\item{then the inelastic $n$-th moments obtained using these distorted data, $M_n^{mod}(Q^2)$, are compared with the original ones $M_n^{in}(Q^2)$ evaluated with no systematic shifts;}

\item{finally the deviations for each data set are summed in quadrature as independent values.}

\end{itemize}

\indent Explicitly, one has
 \be
    \delta_n^D(Q^2) = \frac{1}{M_n^{in}(Q^2)} ~ \sqrt{\sum\limits_i^{N_{S}} 
    \left[ M_n^{mod}(Q^2) - M_n^{in}(Q^2) \right]^2} 
    \label{eq:d_nm_ds1}
 \ee
where $N_{S}$ is the number of available data sets. Then the resulting error is summed in quadrature to $\delta_n^{norm}(Q^2)$.

\indent The second type of error is related to bin centering, interpolation and extrapolation. The systematic uncertainty due to bin centering is estimated as follows:
 \be
    \delta_n^C(Q^2) = \sum_i K_n(x_i,Q^2) ~ w_i(Q^2) ~ \delta^{cent}_{F_2}
    (x_i, Q^2)
 \ee
where according the Nachtmann moment definition and the trapezoidal integration rule one has
 \be
    K_n(x_i, Q^2) & = & \frac{\xi_i^{n + 1}}{x_i^3} \frac{3 + 3(n + 1) r_i + 
    n(n + 2) r_i^2}{(n + 2)(n + 3)} ~ , \nonumber \\
    w_i(Q^2) & = & (x_{i + 1} - x_{i - 1}) / 2 ~ .
 \ee

\indent The relative systematic error of the interpolation is estimated as the possible change of the fitting function slope in the interpolation interval and it is evaluated as a difference in the normalizations at different edges:
 \be
    \delta_S(Q^2) = \left| \frac{1}{N_i} \sum\limits_{i}^{|x_i - x_a| < 
    \Delta} \frac{F_2(x_i, Q^2)}{F_2^{fit}(x_i, Q^2)} - \frac{1}{N_j} 
    \sum\limits_{j}^{|x_j - x_b |< \Delta} \frac{F_2(x_j, Q^2)}{F_2^{fit}(x_j,
    Q^2)} \right| ~ ,
    \label{eq:d_nm_i4}
 \ee
where $N_i$ and $N_j$ are the number of points used to evaluate the sums. Since the relative function $F_2 / F_2^{fit}$ is a very smooth function of $x$ in $DIS$ kinematics, the linear approximation can provide a good estimate on a limited $x$-interval (smaller than $\nabla$). Thus, the error given by Eq.~(\ref{eq:d_nm_i4}) accounts for such a linear mismatch between the fitting function and the data on the interpolation interval. Therefore, the systematic error introduced in the moments by the interpolation procedure can be estimated as
 \be
    \delta_n^I(Q^2) = \delta_S(Q^2) ~ \int\limits_{x_a}^{x_b} dx 
    \frac{\xi^{n + 1}}{x^3} ~ F_2^{fit}(x,Q_0^2) ~ \frac{3 + 3(n + 1) r 
    + n(n + 2) r^2}{(n + 2)(n + 3)} ~ .
    \label{eq:d_nm_i5}
 \ee

\indent The systematic errors obtained by all these procedures are summed in quadrature:
 \be
    \delta_n^P(Q^2) = \sqrt{\left[ \delta_n^D(Q^2) \right]^2 + 
    \left[ \delta_n^C(Q^2) \right]^2 + \left[ \delta_n^I(Q^2) \right]^2} ~ .
    \label{eq:d_nm_i55}
 \ee

\begin{figure}[htb]

\centerline{
\includegraphics[bb=1cm 6cm 20cm 23cm, scale=0.5]{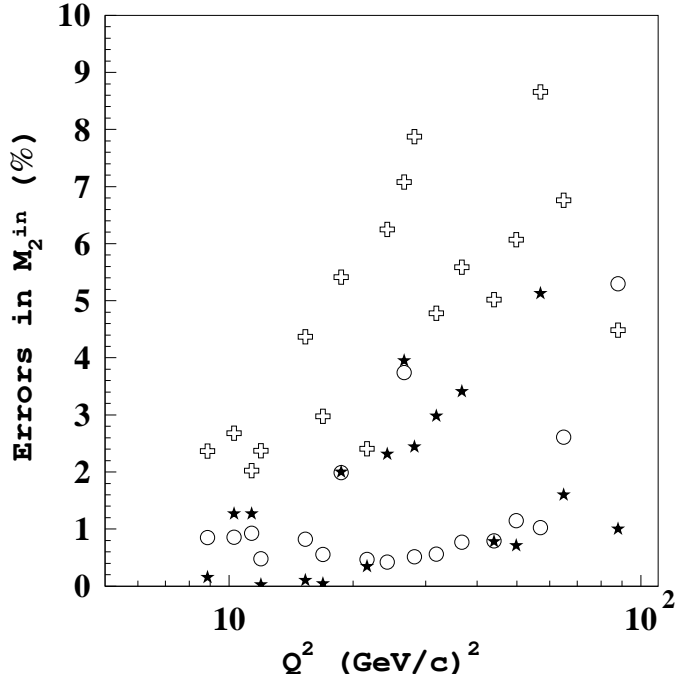}
}

\caption{\label{fig:SepErr} \small \em Errors of the inelastic Nachtmann moment $M_2^{in}(Q^2)$ in percentage: empty circles represent statistical errors; empty crosses show the systematic error obtained in Eq.~(\ref{eq:d_nm_i55}); the difference between inelastic moments extracted using the two different $F_2$ parameterizations of Ref.~\cite{Ricco} and Ref.~\cite{neural} is shown by stars.}

\end{figure}

\indent In order to study the systematic error due to extrapolation in unmeasured regions we have performed a test on a possible dependence upon the functional form of the fitting function by comparing the moments presented here with those obtained using the neural network parameterization of Ref.~\cite{neural}. The difference is expected to be significant only for the second moment $M_2(Q^2)$, due to the presence of a factor $x^{n-2}$ in the integrand of the Nachtmann moments. Any way we have checked the difference for all the moments considered in this work. Such a difference turns out to be smaller than $\delta_n^P(Q^2)$ given by Eq.~(\ref{eq:d_nm_i55}) [see Fig.~\ref{fig:SepErr}] both for $n = 2$ and $n \geq 4$. The difference is added to $\delta^P_n(Q^2)$ in quadrature to evaluate the total systematic error of the $n$-th moment.

\begin{figure}[htb]

\centerline{\includegraphics[scale=0.6]{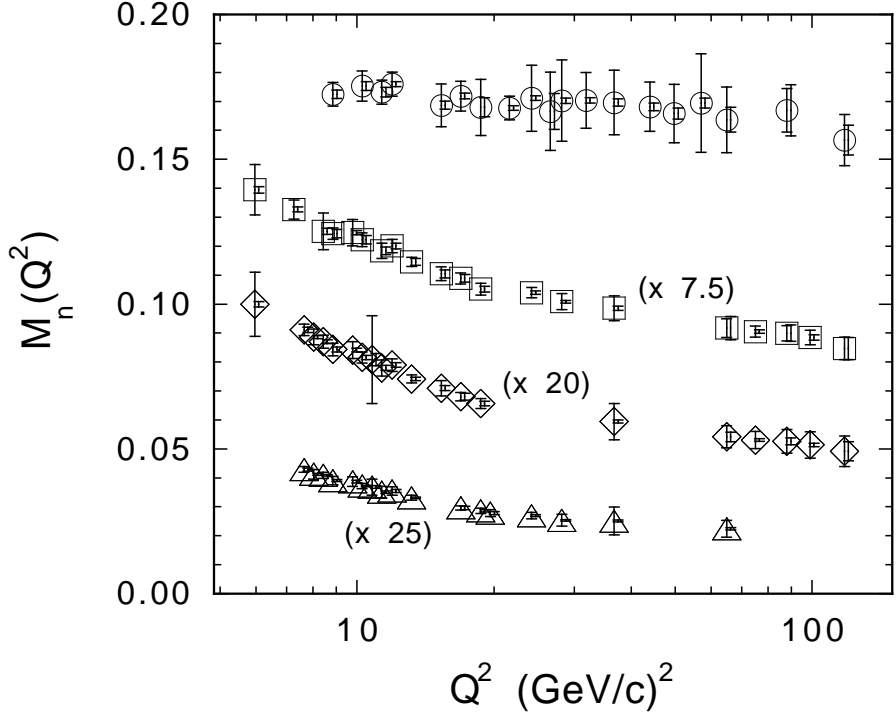}}

\caption{\label{fig:Mn} \small \em Total (inelastic + elastic) Nachtmann moments $M_n(Q^2)$ [Eq.~(\ref{eq:Nachtmann})] extracted from the proton world data in the range $5 < Q^2 < 120 ~ (GeV/c)^2$. Open dots, squares, diamonds and triangles correspond to $n = 2, 4, 6$ and $8$, respectively. Both statistical and systematic errors are reported; the former are those slightly shifted to the right side of the markers.}

\end{figure}

\indent According to Eq.~(\ref{eq:inel+el}) the contribution from the proton elastic peak should be added to the inelastic moments so far obtained. The $Q^2$-behavior of the proton elastic form factor is parameterize as in Ref.~\cite{Drechsel}, modified accordingly to the recent data on $G_E / G_M$ \cite{Jones}. The uncertainty on the form factors is taken to be equal to $3\%$ according to the analysis of Ref.~\cite{Bosted}, and it is added quadratically to both the statistic and the systematic errors. The elastic contribution $M_n^{el}(Q^2)$ turns out to be a quite small correction for $Q^2 > 5 ~ (GeV/c)^2$. Our final results for the total (inelastic + elastic) moments with $n = 2, 4, 6, 8$ including both statistic and systematic errors, are shown in Fig.~\ref{fig:Mn} and reported in Table~\ref{table:nm}. It can cleary be seen that scaling violations are quite small in the second moment $M_2(Q^2)$, while they increase significantly as the order $n$ increases. Consequently, any extraction of $\alpha_s(M_Z)$ from the second moment is expected to be plagued by very large uncertainties, as we have explicitly verified. It is for this reason that in the next Section the $QCD$ analysis of scaling violations will be restricted to $n \geq 4$. Note also that the amount of the direct experimental contribution to $M_n(Q^2)$ reduces between $50\%$ and $70\%$ at large $Q^2$; in particular: at $Q^2 \gsim 75 ~ (GeV/c)^2$ for $n = 4$ and $n =6$, but already at $Q^2 \approx 20 ~ (GeV/c)^2$ for $n = 8$. Finally the systematic uncertainties increase significantly as $Q^2$ increases. 

\indent Before closing this Section, we want to stress two important points:

\begin{itemize}

\item{from Fig.~\ref{fig:Mn} it can clearly be seen that for $n \geq 4$ the $Q^2$-slope of the moments changes significantly around $Q^2 \approx 20 \div 30 ~ (GeV/c)^2$. In our opinion this should be related to the crossing of the $b$-quark threshold, $Q_b^2$, which here after we fix at $Q_b^2 = 25 ~ (GeV/c)^2$. The $Q^2$-region between $\approx 5$ and $\approx 20 ~ (GeV/c)^2$ turns out to be the most sensitive one to scaling violations and, therefore, to the effects of high-order radiative corrections, as already shown in Ref.~\cite{SIM00} (see in particular Fig.~1 of Ref.~\cite{SIM00}). Therefore we consider crucial to include such a region for a precise determination of $\alpha_s(M_Z)$, provided possible power correction effects are taken care of;}

\item{thanks to the direct use of data as well as of parameterizations like the one of Ref.~\cite{neural} our evaluation of the proton Nachtmann moments is independent of any specific choice of the $x$-shape of input parton distributions.}

\end{itemize}

\begin{table}[htb]

\caption{\small \em Total (inelastic + elastic) Nachtmann moments $M_n(Q^2)$ for $n = 2, 4, 6$ and $8$ evaluated in the interval $5 < Q^2 < 120 ~ (GeV/c)^2$. The results, reported together with their statistical and systematic errors, are labeled with an asterisk when the contribution to the integral by the experimental data is between $50\%$ and $70\%$ of the total; all other results are obtained with more than $70\%$ direct contribution from data. Adapted from Ref.~\cite{JLAB}.}

\label{table:nm}

\begin{center}

\scriptsize

\begin{tabular}{|c|c|c|c|c|} \cline{1-5}
$Q^2~[(GeV/c)^2]$ & $M_2(Q^2)$x$10^{-1}$ & $M_4(Q^2)$x$10^{-2}$ & $M_6(Q^2)$x$10^{-3}$ & $M_8(Q^2)$x$10^{-3}$ \\ \cline{1-5}
 5.967 &                               & 1.860 $\pm$ 0.015 $\pm$ 0.116 & 4.995 $\pm$ 0.046 $\pm$ 0.553 &                               \\ \cline{1-5}
 7.268 &                               & 1.768 $\pm$ 0.012 $\pm$ 0.044 &                               &                               \\ \cline{1-5}
 7.646 &                               &                               & 4.552 $\pm$ 0.045 $\pm$ 0.098 & 1.714 $\pm$ 0.021 $\pm$ 0.040 \\ \cline{1-5}
 8.027 &                               &                               & 4.429 $\pm$ 0.027 $\pm$ 0.135 & 1.647 $\pm$ 0.011 $\pm$ 0.064 \\ \cline{1-5}
 8.434 &                               & 1.668 $\pm$ 0.014 $\pm$ 0.084 & 4.350 $\pm$ 0.033 $\pm$ 0.109 & 1.633 $\pm$ 0.013 $\pm$ 0.042 \\ \cline{1-5}
 8.857 & 1.724 $\pm$ 0.015 $\pm$ 0.041 & 1.658 $\pm$ 0.019 $\pm$ 0.027 & 4.215 $\pm$ 0.042 $\pm$ 0.109 & 1.562 $\pm$ 0.014 $\pm$ 0.057 \\ \cline{1-5}
 9.781 &                               & 1.662 $\pm$ 0.010 $\pm$ 0.061 & 4.205 $\pm$ 0.034 $\pm$ 0.146 & 1.546 $\pm$ 0.017 $\pm$ 0.067 \\ \cline{1-5}
10.267 & 1.753 $\pm$ 0.015 $\pm$ 0.052 & 1.629 $\pm$ 0.019 $\pm$ 0.031 & 4.079 $\pm$ 0.035 $\pm$ 0.095 & 1.486 $\pm$ 0.011 $\pm$ 0.037 \\ \cline{1-5}
10.793 &                               &                               & 4.039 $\pm$ 0.106 $\pm$ 0.761 & 1.471 $\pm$ 0.018 $\pm$ 0.110 \\ \cline{1-5}
11.345 & 1.732 $\pm$ 0.016 $\pm$ 0.041 & 1.578 $\pm$ 0.018 $\pm$ 0.035 & 3.896 $\pm$ 0.041 $\pm$ 0.140 & 1.402 $\pm$ 0.018 $\pm$ 0.079 \\ \cline{1-5}
11.939 & 1.759 $\pm$ 0.008 $\pm$ 0.042 & 1.600 $\pm$ 0.013 $\pm$ 0.031 & 3.946 $\pm$ 0.040 $\pm$ 0.111 & 1.418 $\pm$ 0.018 $\pm$ 0.057 \\ \cline{1-5}
13.185 &                               & 1.528 $\pm$ 0.016 $\pm$ 0.021 & 3.705 $\pm$ 0.029 $\pm$ 0.067 & 1.323 $\pm$ 0.009 $\pm$ 0.034 \\ \cline{1-5}
15.310 & 1.686 $\pm$ 0.014 $\pm$ 0.074 & 1.473 $\pm$ 0.019 $\pm$ 0.032 & 3.547 $\pm$ 0.044 $\pm$ 0.133 &                               \\ \cline{1-5}
16.902 & 1.718 $\pm$ 0.010 $\pm$ 0.051 & 1.451 $\pm$ 0.017 $\pm$ 0.025 & 3.401 $\pm$ 0.058 $\pm$ 0.073 & 1.183 $\pm$ 0.025 $\pm$ 0.028 \\ \cline{1-5}
18.697 & 1.679 $\pm$ 0.033 $\pm$ 0.097 & 1.402 $\pm$ 0.013 $\pm$ 0.027 & 3.281 $\pm$ 0.039 $\pm$ 0.088 & 1.144 $\pm$ 0.015 $\pm$ 0.035 \\ \cline{1-5}
19.629 &                               &                               &                               &*1.111 $\pm$ 0.022 $\pm$ 0.047 \\ \cline{1-5}
21.625 & 1.677 $\pm$ 0.008 $\pm$ 0.041 &                               &                               &                               \\ \cline{1-5}
24.192 &*1.711 $\pm$ 0.007 $\pm$ 0.114 & 1.385 $\pm$ 0.008 $\pm$ 0.024 &                               &*1.077 $\pm$ 0.014 $\pm$ 0.043 \\ \cline{1-5}
26.599 &*1.665 $\pm$ 0.062 $\pm$ 0.135 &                               &                               &                               \\ \cline{1-5}
28.192 &*1.702 $\pm$ 0.009 $\pm$ 0.140 & 1.344 $\pm$ 0.007 $\pm$ 0.037 &                               &*1.012 $\pm$ 0.010 $\pm$ 0.081 \\ \cline{1-5}
31.858 &*1.703 $\pm$ 0.010 $\pm$ 0.096 &                               &                               &                               \\ \cline{1-5}
36.750 &*1.696 $\pm$ 0.013 $\pm$ 0.111 & 1.314 $\pm$ 0.009 $\pm$ 0.057 & 2.971 $\pm$ 0.027 $\pm$ 0.313 &*1.003 $\pm$ 0.014 $\pm$ 0.191 \\ \cline{1-5}
44.000 &*1.681 $\pm$ 0.013 $\pm$ 0.085 &                               &                               &                               \\ \cline{1-5}
49.750 &*1.658 $\pm$ 0.019 $\pm$ 0.101 &                               &                               &                               \\ \cline{1-5}
57.000 &*1.694 $\pm$ 0.017 $\pm$ 0.170 &                               &                               &                               \\ \cline{1-5}
64.884 &*1.636 $\pm$ 0.043 $\pm$ 0.114 & 1.222 $\pm$ 0.053 $\pm$ 0.044 & 2.708 $\pm$ 0.082 $\pm$ 0.193 &*0.895 $\pm$ 0.016 $\pm$ 0.116 \\ \cline{1-5}
75.000 &                               &*1.206 $\pm$ 0.008 $\pm$ 0.025 &*2.651 $\pm$ 0.024 $\pm$ 0.150 &                               \\ \cline{1-5}
88.000 &*1.669 $\pm$ 0.088 $\pm$ 0.075 &*1.199 $\pm$ 0.038 $\pm$ 0.035 &*2.630 $\pm$ 0.057 $\pm$ 0.202 &                               \\ \cline{1-5}
99.000 &                               &*1.179 $\pm$ 0.012 $\pm$ 0.034 &*2.568 $\pm$ 0.029 $\pm$ 0.228 &                               \\ \cline{1-5}
117.75 &*1.566 $\pm$ 0.051 $\pm$ 0.088 &*1.128 $\pm$ 0.050 $\pm$ 0.052 &*2.457 $\pm$ 0.162 $\pm$ 0.261 &                               \\ \cline{1-5}

\end{tabular}

\end{center}

\normalsize

\end{table}

\section{$QCD$ analysis of proton moments with $n \geq 4$}
\label{sec:QCD}

\indent The $Q^2$-behavior of the moments $M_n(Q^2)$ can be analyzed adopting the powerful tool of the Operator Product Expansion ($OPE$). Our extracted Nachtmann moments can be expanded as a series of twist terms
 \be
    M_n(Q^2) = \eta_n(Q^2) + \mbox{HT}_n(Q^2)
    \label{eq:OPE}
 \ee
where $\eta_n(Q^2)$ is the leading twist moment of order $n$ and $\mbox{HT}_n(Q^2)$ stands for the total higher-twist contribution, which is power suppressed.

\indent The twist-2 moments $\eta_n(Q^2)$ are generally given by the sum of a singlet and non-singlet ($NS$) terms. The corresponding anomalous dimensions are fixed by $pQCD$ evolution and therefore one is left with a total of three unknown parameters, namely the values of the gluon, singlet and $NS$ quark moments at a given scale $\mu^2$. However, the decoupling in the $pQCD$ evolution of the singlet quark and gluon densities at large $x$ allows to consider a pure $NS$ evolution for $n \geq 4$ (cf.~\cite{Ricco}). This means that, besides the quantity $\alpha_s(\mu^2)$, we have only one twist-2 parameter for $n \geq 4$, namely the value of the twist-2 moment $\eta_n(\mu^2)$ at the chosen scale $\mu^2$. The $NS$ approximation is expected to be quite reasonable for an exploratory study, like the present one. The inclusion of the full singlet and $NS$ evolutions for the moments with $n \geq 4$ is postponed to a future work.

\indent As far as the order of perturbation theory is concerned, the $NLO$ approximation is not reliable for moments with order $n \geq 4$ \cite{SIM00}, which are dominated by kinematical regions corresponding to large values of $x$. Therefore we estimate radiative corrections beyond the $NLO$ according to the soft-gluon resummation ($SGR$) technique developed in Refs.~\cite{SGR,Catani}. We stress that $SGR$, unlike renormalon models, does not introduce any further parameter in the description of the leading twist. With respect to Ref.~\cite{SIM00}, where $SGR$ was considered for the quark coefficient function only, we consistently add in this work the resummation of large $n$-logarithms appearing also in the one-loop and two-loop $NS$ anomalous dimensions\footnote{Such an addition turns out to have a quite small effect ($+ 0.0003$) on our final determination of $\alpha_s(M_Z)$.}. Explicitly, for $n \geq 4$ one has
 \be
    \eta_n(Q^2) & = & A_n ~ \left[ \alpha_s(Q^2) \right]^{\gamma_n^{NS}} \cdot 
    \left\{ {\alpha_s(Q^2) \over 4 \pi} R_n^{NS} \right. \nonumber \\
    & + & \left. \left[ 1 + {\alpha_s(Q^2) \over 4 \pi} \left( 2 
    C_{DIS}^{(NLO)} + \Delta \gamma_{DIS}^{(1, NS)} \right) \right] 
    e^{G_n(Q^2)} \right\} ~ , ~ 
    \label{eq:SGR}
 \ee
where $A_n$ is a constant absorbing in its definition the $n$-th moment of the leading twist at the scale $\mu^2$, and $\gamma_n^{NS}$ is the one-loop $NS$ anomalous dimension. In Eq.~(\ref{eq:SGR}) the quantity $R_n^{NS}$ is given by
 \be
    R_n^{NS} = 2 \left[ C_n^{(NLO)} - C_{DIS}^{(NLO)} - C_{n, LOG}^{(NLO)} 
    \right] + \Delta \gamma_n^{(1, NS)} - \Delta \gamma_{DIS}^{(1, NS)} - 
    \Delta \gamma_{n, LOG}^{(1, NS)}
    \label{eq:RNS}
 \ee
where 
 \be
     \Delta \gamma_n^{(1, NS)} \equiv \gamma_n^{(1, NS)} - {\beta_1 \over 
     \beta_0} \gamma_n^{NS}
     \label{eq:deltagamma}
 \ee
 with $\gamma_n^{(1, NS)}$ being the two-loop $NS$ anomalous dimension, $\beta_0 = 11 - 2 N_f / 3$, $\beta_1 = 102 - 38 N_f / 3$ and $N_f$ the number of active quark flavors at the scale $Q^2$.

\indent In Eq.~({\ref{eq:RNS}) $C_n^{(NLO)}$ is the $NLO$ part of the quark coefficient function, that in the $\overline{MS}$ scheme (which is adopted throughout this paper) reads as
 \be
    C_n^{(NLO)} = C_F \left\{ S_1(n) \left[ S_1(n) + {3 \over 2} - {1 
    \over n(n+1)} \right] - S_2(n) + {3 \over 2n} + {2 \over n+1} + {1 
    \over n^2} - {9 \over 2} \right\}
    \label{eq:Cn}
 \ee
where $C_F \equiv (N_c^2 - 1) / (2 N_c)$ and $S_k(n) \equiv \sum_{j=1}^n 1 /j^k$. For large $n$ (corresponding to the large-$x$ region) the coefficient $C_n^{(NLO)}$ is logarithmically divergent; indeed, since $S_1(n) = \gamma_E + \mbox{log}(n) + O(1/n)$, where $\gamma_E = 0.577216$ is the Euler-Mascheroni constant, and $S_2(n) = \pi^2/6 + O(1/n)$, one gets
 \be
     C_n^{(NLO)} & = & C_{DIS}^{(NLO)} + C_{n, LOG}^{(NLO)} + O(1/n) ~ ,
     \nonumber \\
     C_{DIS}^{(NLO)} & = & C_F \left[ \gamma_E^2 + {3 \over 2} \gamma_E - 
     {9 \over 2} - {\pi^2 \over 6} \right] ~ , \nonumber \\
     C_{n, LOG}^{(NLO)} & = & C_F ~ \mbox{ln}(n) \left[ \mbox{ln}(n) + 
     2\gamma_E + {3 \over 2} \right] ~ .
     \label{eq:Cn_exp}
 \ee
In case of the quantity $\Delta \gamma_n^{(1, NS)}$ one obtains
 \be
     \label{eq:gammas}
     \Delta \gamma_n^{(1, NS)} & = & \Delta \gamma_{DIS}^{(1, NS)} + \Delta 
     \gamma_{n, LOG}^{(1, NS)} + O(1 / n) ~ , \nonumber \\
     \Delta \gamma_{DIS}^{(1, NS)} & = & {C_F \over \beta_0} \left\{ C_F 
     \left[ 2\pi^2 + 32 \tilde{S}(\infty) - 4 S_3(\infty) - {3 \over 2} 
     \right] \right. \nonumber \\
      & + & \left. C_A \left[ -{22 \over 9} \pi^2 - 16 \tilde{S}(\infty) - 
      {17 \over 6} \right] \right. \nonumber \\
      & + & \left. N_f \left[ {4 \pi^2 \over 9} + {1 \over 3} \right] + 
      \gamma_E \left( 8 K - 4 {\beta_1 \over \beta_0} \right) + 3 {\beta_1 
      \over \beta_0} \right\} ~ , \nonumber \\
      \Delta \gamma_{n, LOG}^{(1, NS)} & = & {C_F \over \beta_0} \left[ 8 K -
      4 {\beta_1 \over \beta_0} \right] \mbox{ln}(n) ~ ,
 \ee
where $C_A = N_c$, $\tilde{S}(\infty) = \sum_{j = 1}^{\infty} (-)^j S_1(j) / j^2 = -0.751286$, $S_3(\infty) = 1.202057$ and $K = C_A ~ (67/18 - \pi^2 / 6) - 5 N_f / 9$.

\indent In Eq.~(\ref{eq:SGR}) the function $G_n(Q^2)$ is the key quantity of the soft gluon resummation. At $NLL$ accuracy it reads as 
 \be
    G_n(Q^2) = \mbox{ln}(n) ~ G_1(\lambda_n) + G_2(\lambda_n) + O\left[ 
    \alpha_s^k \mbox{ln}^{k-1}(n) \right] ~ ,
   \label{eq:Gn}
\ee
where $\lambda_n \equiv \beta_0 ~ \alpha_s(Q^2) ~ \mbox{ln}(n) / 4\pi$ and
 \be
    G_1(\lambda) & = & C_F {4 \over \beta_0 \lambda} \left[ \lambda + (1 
    -  \lambda) \mbox{ln}(1 - \lambda) \right] ~ , \nonumber \\[3mm]
    G_2(\lambda) & = & - C_F {4 \gamma_E + 3 \over \beta_0} \mbox{ln}(1 
    - \lambda) - C_F {8 K \over \beta_0^2} \mbox{ln}(1 - \lambda) 
    \nonumber \\
    & & + C_F {4  \beta_1 \over \beta_0^3} \mbox{ln}(1 - \lambda) \left[ 1 + 
    {1 \over 2} \mbox{ln}(1 - \lambda) \right] ~ ,
    \label{eq:G1G2}
 \ee
with $K = N_c ~ (67/18 - \pi^2 / 6) - 5 N_f / 9$. Note that the function $G_2(\lambda)$ is divergent for $\lambda \to 1$; this means that at large $n$ (i.e. large $x$) the soft gluon resummation cannot be extended to arbitrarily low values of $Q^2$. Moreover, the sub-leading terms $O[\alpha_s^k \mbox{ln}^{k-1}(n)]$ become more and more important as $\lambda_n$ increases. Therefore, for a safe use of the $SGR$ technique at $NLL$ accuracy it is essential to check that $\lambda_n$ is small enough. Later on, we will impose the condition that the $Q^2$-ranges of our analysis should correspond to the cut $\lambda_n \lsim 0.3$; this appears to be a safe upper limit at least to avoid significant effects from $NNLL$ terms in Eq.~(\ref{eq:Gn}), as we have checked using the $NNLL$ order expression for the function $G_n(Q^2)$ reported in Ref.~\cite{Vogt}. It is straightforward to see that in the limit $\lambda_n << 1$ one has $G_n(Q^2) \to \alpha_s(Q^2) ~ [ 2 ~ C_{n, LOG}^{(NLO)} + \Delta \gamma_{n, LOG}^{(1, NS)} ] / 4 \pi$, so that Eq. (\ref{eq:SGR}) reduces to the well-known $NLO$ approximation. This implies that adopting the two-loop approximation for the running coupling constant $\alpha_s(Q^2)$ the twist-2 expression (\ref{eq:SGR}) contains all the $NLO$ effects and the resummation of large $n$-logarithms beyond the $NLO$. 

\indent To sum up, the leading twist (\ref{eq:SGR}) contains two unknown parameters: the coefficient $A_n$ and the value of the coupling constant $\alpha_s(\mu^2)$ at the scale $\mu^2$. Alternatively, as another parameterization for $\alpha_s(Q^2)$, the so-called $QCD$ scale parameter $\Lambda$ is introduced in a such a way that at two-loops the running of the coupling constant $\alpha_s(Q^2)$ can be written as
 \be
    \alpha_s(Q^2) = {4 \pi \over \beta_0 ~ \mbox{ln}(Q^2 / \Lambda^2)} \left\{ 
    1 - {\beta_1 \over \beta_0} {\mbox{ln}[ \mbox{ln}(Q^2 / \Lambda^2) ] \over 
    \mbox{ln}( Q^2 / \Lambda^2)} \right\} ~ .
    \label{eq:alphas}
 \ee
Actually the parameter $\Lambda$ depends on the renormalization scheme (the $\overline{MS}$ one in this paper) and on the number of active quark flavors. Therefore, more properly, in Eq.~(\ref{eq:alphas}) $\Lambda$ should be labelled as $\Lambda_{\overline{MS}}^{(N_f)}$. The constraints of continuity of $\alpha_s(Q^2)$ at the various quark thresholds impose relations between the parameters $\Lambda_{\overline{MS}}^{(N_f)}$ for the various values of $N_f$, leaving only one of them as a free parameter to be determined from the data.

\indent As far as power corrections are concerned, several higher-twist ($HT$) operators exist and mix under the renormalization-group equations. Such mixings are rather involved and the number of mixing operators increases rapidly with the order $n$ of the moment. A complete calculation of the $HT$ anomalous dimensions for $n > 2$ is not yet available, and therefore a phenomenological ansatz is usually adopted for $\mbox{HT}_n(Q^2)$ in the literature (cf.~Refs.~\cite{Ricco,SIM00,JLAB}), namely
 \be
    \mbox{HT}_n(Q^2) = {a_n^{(4)} \over Q^2} \left[ \alpha_s(Q^2) 
    \right]^{\gamma_n^{(4)}} + {a_n^{(6)} \over Q^4} \left[ \alpha_s(Q^2) 
    \right]^{\gamma_n^{(6)}} + \mbox{...} ~,
    \label{eq:HT}
 \ee
where the logarithmic pQCD evolution of the twist-$\tau$ contribution is accounted for by the term $[\alpha_s(Q^2)]^{\gamma_n^{(\tau)}}$ with an {\em effective} anomalous dimension $\gamma_n^{(\tau)}$ and the parameter $a_n^{(\tau)}$ represents the overall strength of the twist-$\tau$ term. Finally, in Eq.~(\ref{eq:HT}) the ellipses stand for contributions of twists higher than $\tau = 6$.

\indent The twist analysis of Ref.~\cite{JLAB} includes the new precise data set obtained with the detector $CLAS$ at Jefferson Lab, which dominates the $Q^2$-region below $\approx 5 ~ (GeV/c)^2$. In Ref.~\cite{JLAB}, where all the $HT$ parameters appearing in Eq.~(\ref{eq:HT}) were determined from data fitting, it was found that power corrections are relevant in the few $(GeV/c)^2$ region, and for $Q^2 \gsim 1 ~ (GeV/c)^2$ they can be described fairly well by the first two terms of the $HT$ expansion (\ref{eq:HT}). The $Q^2$-behavior of the ratio between the total $HT$ term (\ref{eq:HT}) and the total moments $M_n(Q^2)$, as determined in Ref.~\cite{JLAB}, is reported in Fig.~\ref{fig:HT} for $Q^2 > 5 ~ (GeV/c)^2$.

\begin{figure}[htb]

\centerline{\includegraphics[scale=0.6]{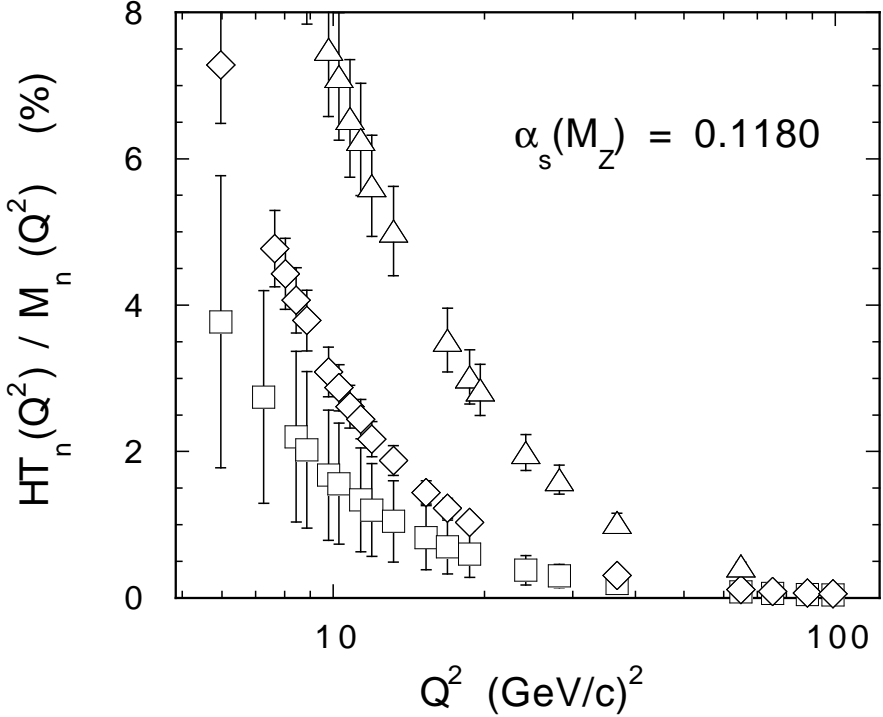}}

\caption{\label{fig:HT} \small \em Ratio in percentage between the total $HT$ contribution $HT_n(Q^2)$ [Eq.~(\ref{eq:HT})] and the moments $M_n(Q^2)$ for $Q^2 > 5 ~ (GeV/c)^2$. The $HT$ contribution has been obtained in Ref.~\cite{JLAB} using proton world data in the $Q^2$-range $1 \lsim Q^2 ~ (GeV/c)^2 \lsim 100$. The value of $\alpha_s(M_Z)$ is kept fixed at $\alpha_s(M_Z) = 0.1180$. Open squares, diamonds and triangles correspond to $n = 4, 6$ and $8$, respectively. Error bars represent the uncertainties resulting from the fitting procedure of Ref.~\cite{JLAB}. Adapted from Ref.~\cite{JLAB}.}

\end{figure}

\indent In Ref.~\cite{JLAB} the value of $\alpha_s(M_Z)$ was not determined from the data, but it was kept fixed close to the world average, namely $\alpha_s(M_Z) = 0.1180$. The reason for that is the occurrence of an (anti)correlation between the value of $\alpha_s(M_Z)$ and the total $HT$ strength. Such a correlation does not allow to extract $\alpha_s(M_Z)$ and the $HT$ simultaneously from our proton data with good precision. This point was already addressed in Ref.~\cite{SIM00}, where the impact of the specific value of $\alpha_s(M_Z)$ on the extracted higher twists was investigated. There it was shown that the strength of the total higher-twist decreases as $\alpha_s(M_Z)$ increases without changing the statistical significance of the fit; indeed, the calculated $\chi^2$ turned out to be almost insensitive to the specific combination of the values of $\alpha_s(M_Z)$ and of the total $HT$ strength.

\indent The previous considerations clearly suggest that for the extraction of $\alpha_s(M_Z)$ from the moments $M_n(Q^2)$ a pure leading-twist analysis is necessary, but the lower limit $Q_{min}^2$ of the range of the analysis should  be chosen in such a way that perturbative effects are maximized for each order $n$ separately, while $HT$ effects can be still considered under control. Both goals can be achieved as follows:

\begin{itemize}

\item{since we work within the $NLL$ approximation for the resummation of large $n$-logarithms, we have to impose the condition that $NNLL$ effects are negligible, which means that we should keep the parameter $\lambda_n = \beta_0 ~ \alpha_s(Q^2) ~ \mbox{ln}(n) / 4\pi$, appearing in Eq.~(\ref{eq:Gn}), small enough. Effects of the $NNLL$ order on the $SGR$ quantity $G_n(Q^2)$ have been calculated in Ref.~\cite{Vogt} and we have checked that the cut $\lambda_n \lsim 0.3$ makes those $NNLL$ effects negligible for $n = 4, 6$ and $8$. We point out however that our present estimate of $NNLL$ effects is not complete. Effects of subleading logs have been recently investigated in Ref.~\cite{GR03}, where an upper bound $\lambda_n \simeq 0.4$ was assumed at $NNLL$ accuracy. At $\alpha_s(M_Z) = 0.1180$ our upper bound $\lambda_n \simeq 0.3$ corresponds to $Q^2 \approx 5, 7$ and $11 ~ (GeV/c)^2$ for $n = 4, 6$ and $8$, respectively;}

\item{from Table~\ref{table:nm} the average value of the systematic errors on our moments in the $Q^2$-range from $\approx 5$ to $\approx 20 ~ (GeV/c)^2$ turns out to be $\approx 3\%$ at $n = 4, 6$ and $\approx 4\%$ at $n = 8$. Looking now at Fig.~\ref{fig:HT} it can be seen that such percentage values are reached by the total $HT$ contribution when $Q^2 \approx 9, 10$ and $16 ~ (GeV/c)^2$ for $n = 4, 6$ and $8$, respectively.}

\end{itemize}

\indent Therefore, for the lower limit $Q_{min}^2$ we will consider the following ranges of values in units of $[(GeV/c)^2]$: $5 \lsim Q_{min}^2 \lsim 9$ for $n = 4$; $7 \lsim Q_{min}^2 \lsim 10$ for $n = 6$; and $11 \lsim Q_{min}^2 \lsim 16$ for $n = 8$. At the same time, in order to be as much conservative as possible, the total $HT$ strength shown in Fig.~\ref{fig:HT} is added (in quadrature) to the systematic error of our moments. Finally, the upper limit of the $Q^2$-range of our analysis is taken to be equal to $120 ~ (GeV/c)^2$ for each $n$ according to the data set of Table~\ref{table:nm}.

\indent Using a $\chi^2$-minimization procedure, we have fitted the moments $M_n(Q^2)$ neglecting the $HT$ contribution $HT_n(Q^2)$ in Eq.~(\ref{eq:OPE}) and adopting the $SGR$ formula (\ref{eq:SGR}) for the leading-twist term $\eta_n(Q^2)$. The results obtained at $n = 4, 6$ and $8$ for various values of $Q_{min}^2$ are reported in Table~\ref{table:Mn}. There, the columns labelled "stat." represent the uncertainties due to the fitting procedure  and driven by the statistical errors of our moments. The columns labelled "syst." represent our estimate of the systematic uncertainties on the extracted values of $\alpha_s(M_Z)$, obtained simply by adding (subtracting) to the central values of the moments their own systematical errors and by repeating the $\chi^2$-minimization procedure. In this way an upper (lower) value of $\alpha_s(M_Z)$ is generated by the systematic errors of the moments.

\begin{table}[htb]

\caption{\small \em Values of $\alpha_s(M_Z)$ extracted from a pure leading-twist analysis of the moments $M_n(Q^2)$, using the $SGR$ formula (\ref{eq:SGR}), for various values of the lower limit $Q_{min}^2$ of the analysis range, as described in the text. The values of $Q_{min}^2$ are in $(GeV/c)^2$, while the upper limit is fixed at $Q_{max}^2 = 120 ~ (GeV/c)^2$. The columns labelled "stat." and "syst." represent the uncertainties due to the fitting procedure and to the systematic errors of the moments $M_n(Q^2)$, respectively (see text).}

\label{table:Mn}

\begin{center}

\small

\begin{tabular}{|c|} \hline
$\mbox{n = 4}$ \\ \hline

\end{tabular}

\begin{tabular}{||c|c||c|c|c||} \cline{1-5}
$Q_{min}^2$ & $\mbox{$N^o$ of data}$ & $\alpha_s(M_Z)$ & $\mbox{stat.}$ & $\mbox{syst.}$ \\ \cline{1-5}
5 & 20 & 0.1198 & 0.0027 & 0.0009 \\ \cline{1-5}
7 & 19 & 0.1191 & 0.0030 & 0.0005 \\ \cline{1-5}
8 & 18 & 0.1179 & 0.0035 & 0.0009 \\ \cline{1-5}
9 & 16 & 0.1183 & 0.0037 & 0.0006 \\ \cline{1-5}

\end{tabular}

\vspace{0.25cm}

\begin{tabular}{|c|} \hline
$\mbox{n = 6}$ \\ \hline

\end{tabular}

\begin{tabular}{||c|c||c|c|c||} \cline{1-5}
$Q_{min}^2$ & $\mbox{$N^o$ of data}$ & $\alpha_s(M_Z)$ & $\mbox{stat.}$ & $\mbox{syst.}$ \\ \cline{1-5}
 7 & 19 & 0.1197 & 0.0022 & 0.0011 \\ \cline{1-5}
 8 & 18 & 0.1194 & 0.0022 & 0.0013 \\ \cline{1-5}
 9 & 15 & 0.1181 & 0.0027 & 0.0025 \\ \cline{1-5}
10 & 14 & 0.1169 & 0.0030 & 0.0032 \\ \cline{1-5}

\end{tabular}

\vspace{0.25cm}

\begin{tabular}{|c|} \hline
$\mbox{n = 8}$ \\ \hline

\end{tabular}

\begin{tabular}{||c|c||c|c|c||} \cline{1-5}
$Q_{min}^2$ & $\mbox{$N^o$ of data}$ & $\alpha_s(M_Z)$ & $\mbox{stat.}$ & $\mbox{syst.}$ \\ \cline{1-5}
11 & 10 & 0.1214 & 0.0030 & 0.0050 \\ \cline{1-5}
12 &  9 & 0.1213 & 0.0027 & 0.0078 \\ \cline{1-5}
13 &  8 & 0.1196 & 0.0034 & 0.0088 \\ \cline{1-5}
16 &  7 & 0.1085 & 0.0039 & 0.0105 \\ \cline{1-5}

\end{tabular}

\normalsize

\end{center}

\end{table}

\indent From Table~\ref{table:Mn} it can be seen that for $n = 4$ and $n = 6$ all the determinations of $\alpha_s(M_Z)$ are consistent with each other within the statistical error. The systematic uncertainties are quite limited and not larger than the statistical ones. The same does not hold in case of $n = 8$. Indeed, using $Q_{min}^2 = 11, 12$ and $13 ~ (GeV/c)^2$ we still extract values of $\alpha_s(M_Z)$ almost consistent with each other, but larger than those obtained at $n = 4, 6$. Moreover, for $Q_{min}^2 = 16 ~ (GeV/c)^2$ the determination of $\alpha_s(M_Z)$ drops down to a very low value. Finally, the systematic uncertainties become quite large at $n = 8$. Note that: ~ i) the number of data points is the smallest one at $n = 8$; ~ ii) the amount of the direct experimental contribution to $M_8(Q^2)$ reduces between $50\%$ and $70\%$ already at $Q^2 \approx 20 ~ (GeV/c)^2$ with large systematic uncertainties (see Table~\ref{table:nm}). Thus, the results obtained at $n = 8$ may be explained, at least partially, by the limited quality of the data, but they might also represent the signal for the need of including higher-order perturbative effects.

\indent For each moment we have separately calculated the (weighted) average of the extracted values of $\alpha_s(M_Z)$ over the various values of $Q_{min}^2$. The results are collected in Table~\ref{table:avg}. The central values for $n = 4, 6$ are in good agreement with each other and, surprisingly, also the one at $n = 8$. However, we do not consider significant such a coincidence because of the spread of the results reported in Table~\ref{table:Mn} at $n = 8$. Then, we have calculated the (weighted) average of the results obtained for the various values of $n$, and we have reported them again in Table~\ref{table:avg}.

\begin{table}[htb]

\caption{\small \em Weighted averages of the values of $\alpha_s(M_Z)$ reported in Table~\protect\ref{table:Mn} for each moment $M_n(Q^2)$ separately. The last two rows represent the final (weighted) averages of $\alpha_s(M_Z)$ taken over the various moments considered.}

\label{table:avg}

\begin{center}

\small

\begin{tabular}{||c||c|c|c||} \cline{1-4}
$\mbox{moment}$ & $\alpha_s(M_Z)$ & $\mbox{stat.}$ & $\mbox{syst.}$ \\ \cline{1-4}
$M_4$ & 0.1189 & 0.0016 & 0.0007 \\ \cline{1-4}
$M_6$ & 0.1188 & 0.0012 & 0.0018 \\ \cline{1-4}
$M_8$ & 0.1189 & 0.0016 & 0.0077 \\ \cline{1-4} \\ \cline{1-4}
$\mbox{average}$ & 0.1189 & 0.0008 & 0.0031 \\ \cline{1-4}
$M_4 + M_6$ & 0.1188 & 0.0010 & 0.0014 \\ \cline{1-4}

\end{tabular}

\normalsize

\end{center}

\end{table}

\indent As our final estimate of $\alpha_s(M_Z)$ we exclude the determination at $n = 8$ and consider only the (weighted) average of the results at $n = 4$ and $n = 6$, leading to
 \be
    \alpha_s(M_Z) = 0.1188 \pm 0.0010 ~ (stat.) \pm 0.0014 ~ (syst.)
    \label{eq:alphas_MZ}
 \ee
We have checked that a change in the $b$-quark threshold $Q_b^2$ between $\approx 20$ and $\approx 30 ~ (GeV/c)^2$ introduces an uncertainty of $\approx \pm 0.0005$, which can be neglected in comparison with the systematic error reported in Eq.~(\ref{eq:alphas_MZ}). It should be mentioned that the uncertainty in the extraction of $\alpha_s(M_Z)$ may be affected also by the renormalization scale dependence of Eq.~(\ref{eq:SGR}). Such a dependence is known in literature \cite{Catani} and it is a common practice to investigate its impact on the extracted value of $\alpha_s(M_Z)$, obtaining in this way an estimate of higher-order perturbative effects. However, due to the exploratory nature of our present work, we postpone such a (mandatory) investigation to a future work.

\indent In order to disentangle the effects of the soft-gluon resummation we have repeated our calculations adopting the $NLO$ approximation for the leading twist $\eta_n(Q^2)$. The results are reported in the Tables~\ref{table:Mn_NLO}-\ref{table:avg_NLO}. It can be seen that our $NLO$ determination of $\alpha_s(M_Z)$ exhibit features similar to the ones observed for the $SGR$ results of Tables~\ref{table:Mn}-\ref{table:avg}, including the disturbing strong dependence on $Q_{min}^2$ at $n = 8$, but with an important difference: the determinations of $\alpha_s(M_Z)$ for the various values of $n$ are not consistent with each other and a remarkable dependence on the order of the moment is clearly visible in Table~\ref{table:avg_NLO} and in Fig.~\ref{fig:as}, where the results obtained at $NLO$ and within the $SGR$ are compared directly.

\begin{table}[htb]

\caption{\small \em The same as in Table~\protect\ref{table:Mn}, but adopting the $NLO$ approximation for the leading twist.}

\label{table:Mn_NLO}

\begin{center}

\small

\begin{tabular}{|c|} \hline
$\mbox{n = 4}$ \\ \hline

\end{tabular}

\begin{tabular}{||c|c||c|c|c||} \cline{1-5}
$Q_{min}^2$ & $\mbox{$N^o$ of data}$ & $\alpha_s(M_Z)$ & $\mbox{stat.}$ & $\mbox{syst.}$ \\ \cline{1-5}
5 & 20 & 0.1201 & 0.0027 & 0.0008 \\ \cline{1-5}
7 & 19 & 0.1194 & 0.0029 & 0.0004 \\ \cline{1-5}
8 & 18 & 0.1182 & 0.0034 & 0.0004 \\ \cline{1-5}
9 & 16 & 0.1186 & 0.0037 & 0.0006 \\ \cline{1-5}

\end{tabular}

\vspace{0.25cm}

\begin{tabular}{|c|} \hline
$\mbox{n = 6}$ \\ \hline

\end{tabular}

\begin{tabular}{||c|c||c|c|c||} \cline{1-5}
$Q_{min}^2$ & $\mbox{$N^o$ of data}$ & $\alpha_s(M_Z)$ & $\mbox{stat.}$ & $\mbox{syst.}$ \\ \cline{1-5}
 7 & 19 & 0.1232 & 0.0024 & 0.0013 \\ \cline{1-5}
 8 & 18 & 0.1228 & 0.0025 & 0.0014 \\ \cline{1-5}
 9 & 15 & 0.1214 & 0.0029 & 0.0028 \\ \cline{1-5}
10 & 14 & 0.1200 & 0.0032 & 0.0035 \\ \cline{1-5}

\end{tabular}

\vspace{0.25cm}

\begin{tabular}{|c|} \hline
$\mbox{n = 8}$ \\ \hline

\end{tabular}

\begin{tabular}{||c|c||c|c|c||} \cline{1-5}
$Q_{min}^2$ & $\mbox{$N^o$ of data}$ & $\alpha_s(M_Z)$ & $\mbox{stat.}$ & $\mbox{syst.}$ \\ \cline{1-5}
11 & 10 & 0.1276 & 0.0030 & 0.0060 \\ \cline{1-5}
12 &  9 & 0.1273 & 0.0034 & 0.0091 \\ \cline{1-5}
13 &  8 & 0.1154 & 0.0033 & 0.0097 \\ \cline{1-5}
16 &  7 & 0.1125 & 0.0066 & 0.0110 \\ \cline{1-5}

\end{tabular}

\end{center}

\normalsize

\end{table}

\begin{table}[htb]

\caption{\small \em The same as in Table~\protect\ref{table:avg}, but in case of the $NLO$ approximation for the leading twist.}

\label{table:avg_NLO}

\begin{center}

\small

\begin{tabular}{||c||c|c|c||} \cline{1-4}
$\mbox{moment}$ & $\alpha_s(M_Z)$ & $\mbox{stat.}$ & $\mbox{syst.}$ \\ \cline{1-4}
$M_4$ & 0.1192 & 0.0016 & 0.0007 \\ \cline{1-4}
$M_6$ & 0.1221 & 0.0013 & 0.0020 \\ \cline{1-4}
$M_8$ & 0.1258 & 0.0018 & 0.0083 \\ \cline{1-4} \\ \cline{1-4}
$\mbox{average}$ & 0.1221 & 0.0009 & 0.0031 \\ \cline{1-4}
$M_4 + M_6$ & 0.1209 & 0.0010 & 0.0015 \\ \cline{1-4}

\end{tabular}

\normalsize

\end{center}

\end{table}

\begin{figure}[htb]

\centerline{\includegraphics[scale=0.6]{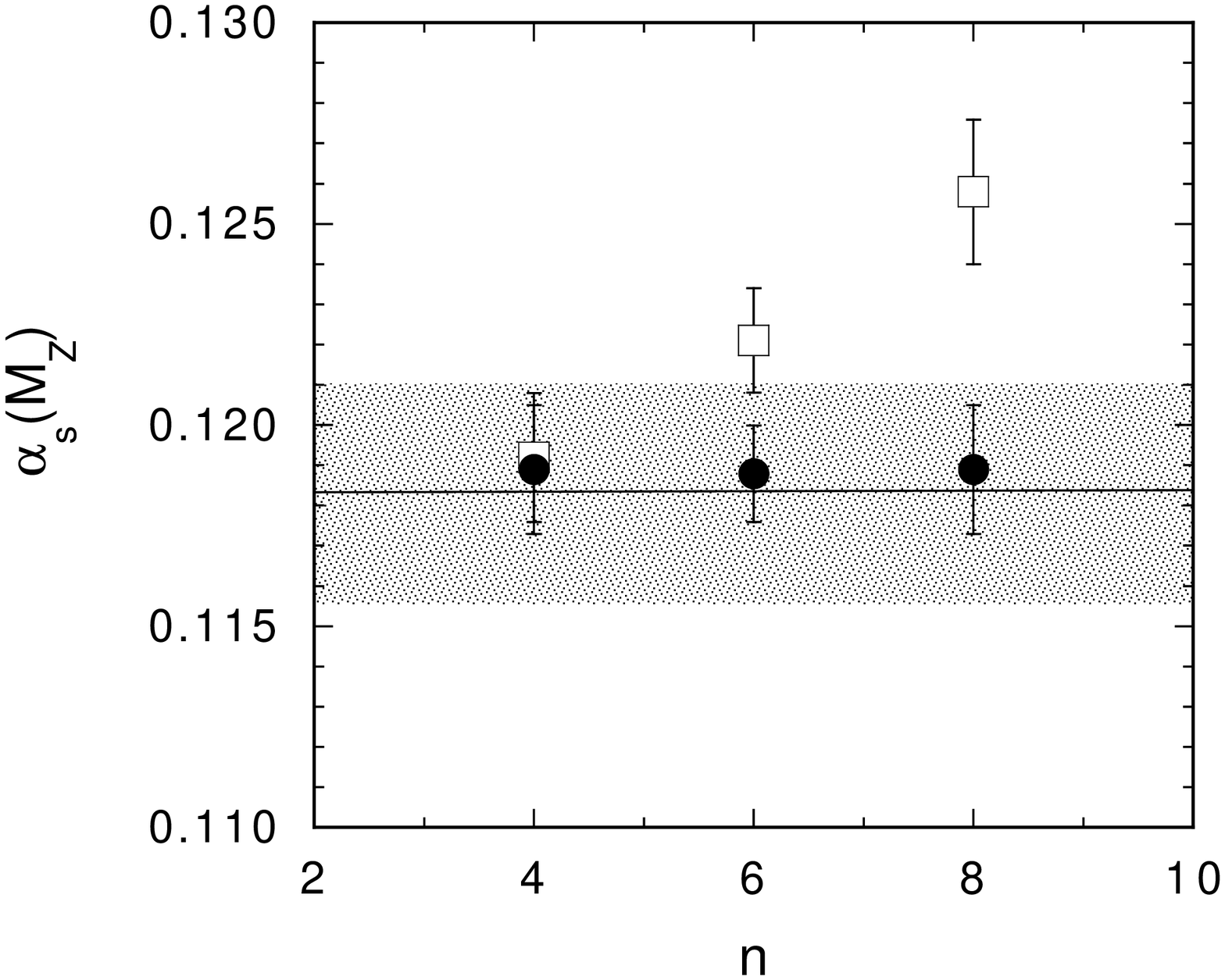}}

\caption{\label{fig:as} \small \em Values of $\alpha_s(M_Z)$ extracted in this work from the moments $M_n(Q^2)$ at $n = 4, 6$ and $8$. Full dots and open squares correspond to the results obtained using the $SGR$ formula (\ref{eq:SGR}) and within the $NLO$ approximation for the leading twist, respectively. The vertical bars represent only the statistical uncertainties due to the fitting procedure. Note that the systematic errors are of the same order of magnitude as the statistical ones for $n = 4$ and $n =6$, but much larger for $n = 8$ (see Tables~\ref{table:avg} and \ref{table:avg_NLO}). The solid line and the shaded area are the world average value and its global uncertainty, as given in Ref.~\cite{update}.}

\end{figure}

\indent In Fig.~\ref{fig:comp} our final determination of $\alpha_s(M_Z)$ [see Eq.~(\ref{eq:alphas_MZ})] is compared with the total and $DIS$ world averages, taken from Ref.~\cite{update}. It can be seen that the comparison is positive and our central value agrees very well with the world averages within the quoted uncertainties. In the same figure we have included also the results of the recent $NNLO$ analyses of Refs.~\cite{SY01} and \cite{Alekhin}. The latter appears to be significantly lower than our determination (\ref{eq:alphas_MZ}), while the result of Ref.~\cite{SY01} is still consistent with our result but at the limits of the quoted uncertainties. We expect that an inclusion of all the $NNLO$ effects with a consistent resummation of large $n$-logarithms at $NNLL$ accuracy will bring our result closer to the one of Ref.~\cite{SY01} and possibly also to the findings of Ref.~\cite{Alekhin}. We point out however that the result of Ref.~\cite{Alekhin} has been obtained with a procedure which depends on the specific choice of the input parton distributions, while our procedure and the one of Ref.~\cite{SY01} are totally free from such a dependence. 

\begin{figure}[htb]

\vspace{0.25cm}

\centerline{\includegraphics[scale=0.6]{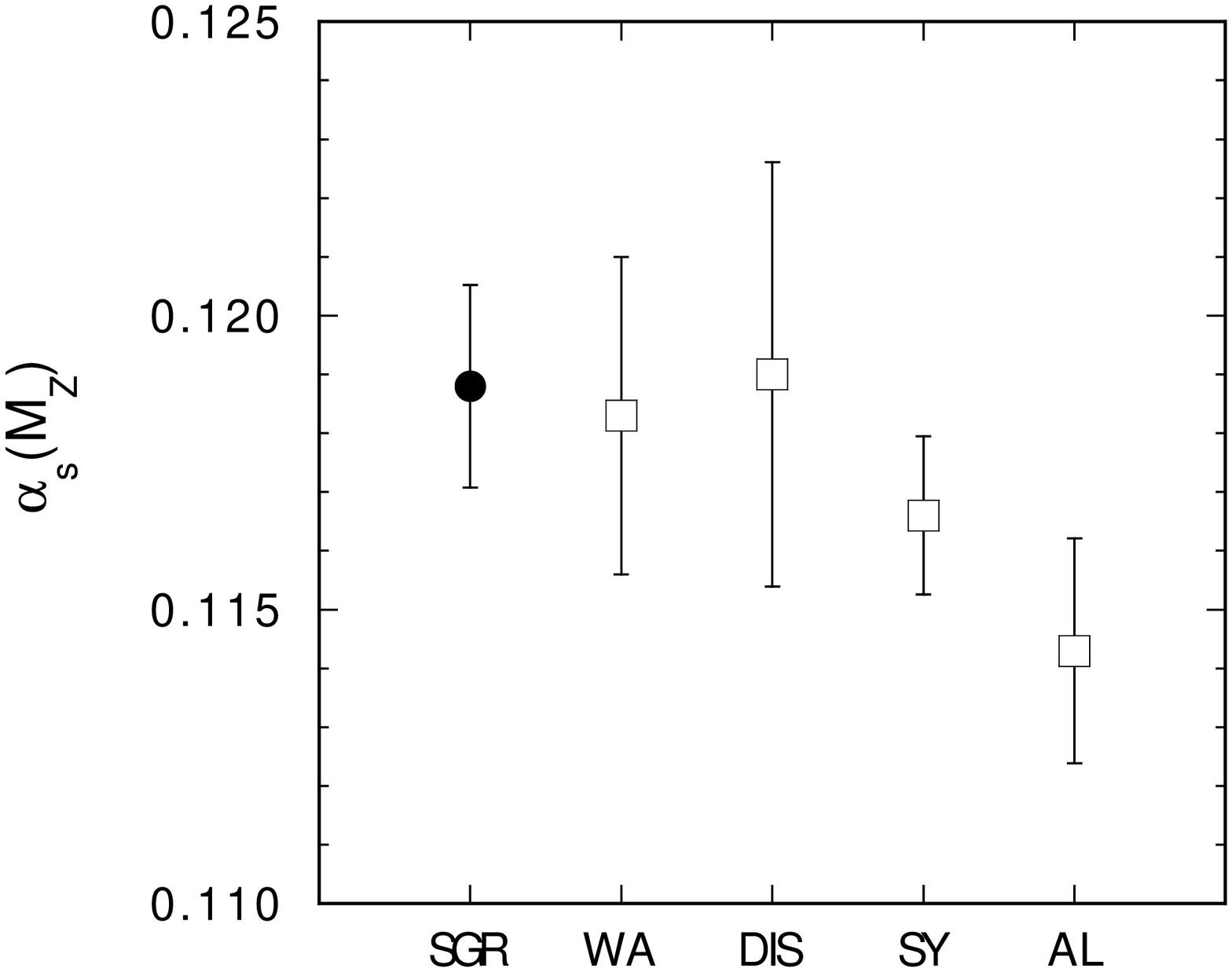}}

\caption{\label{fig:comp} \small \em Comparison of our extracted value of $\alpha_s(M_Z)$ (full dot) with some existing determinations (open squares). Our final value is labelled by $SGR$, which reminds that it is obtained including soft-gluon resummation effects. Statistical and systematic uncertainties are added in  quadrature. The results labelled $WA$ and $DIS$ correspond to the total and the $DIS$ world averages from Ref.~\cite{update}, while those labelled $SY$ and $AL$ come from the recent $NNLO$ analyses of $DIS$ data of Refs.~\cite{SY01} and \cite{Alekhin}, respectively. Experimental and theoretical uncertainties are added in quadrature.}

\end{figure}

\indent The results presented in our exploratory study show that the new approach appears to work quite well and it may represent an interesting tool for providing a precise determination of $\alpha_s(M_Z)$ from $DIS$ data. Clearly our approach should be improved both from the experimental and the theoretical sides.

\indent As far as data are concerned, we need a more precise determination of the Nachtmann moments in the $Q^2$ range from $\approx 5$ to $\approx 20 ~ (GeV/c)^2$, where the largest sensitivity to scaling violations is expected. In this respect we point out that thanks to its large acceptance the $CLAS$ detector at Jefferson Lab has already demonstrated \cite{JLAB} its capability to determine the cross section in a wide two-dimensional range of values of $x$ and $Q^2$, allowing a direct integration of the data at fixed $Q^2$ over the whole significant $x$-range for the determination of the Nachtmann moments with $n > 2$. Presently with the $6 ~ GeV$ electron beam, operating at Jefferson Lab, the upper limit in $Q^2$ is around $5 ~ (GeV/c)^2$. The proposed upgrade to $12 ~ GeV$ electron beam \cite{upgrade} will make it possible to extend the range of $Q^2$ up to $\approx 15 ~ (GeV/c)^2$. In our opinion this will allow a drastic improvement in the quality of the {\em experimental} Nachtmann moments at $Q^2 > 5 ~ (GeV/c)^2$ for $n > 2$.

\indent From the theoretical point of view the main improvements to be developed in our procedure are: ~ i) the consideration of more Nachtmann moments in our analysis in order to check the stability of the extracted values of $\alpha_s(M_Z)$ against the order of the moment; ~ ii) the inclusion of all the $NNLO$ effects with a consistent resummation of large $n$-logarithms at $NNLL$ accuracy; and ~ iii) the consideration of the full singlet and $NS$ evolutions for the moments with $n > 2$.

\section{Conclusions}
\label{sec:conclusions}

\indent An exploratory study for a new determination of the strong coupling constant $\alpha_s(M_Z)$ from existing world data on the structure function $F_2$ of the proton in the $Q^2$-range $5 \lsim Q^2 ~ (GeV/c)^2 \lsim 120$ has been presented. The main features of our approach are: ~ 1) the use of low-order Nachtmann moments evaluated with a direct contribution from data larger than $70\%$ of the total; ~ 2) the inclusion of high-order perturbative effects through the soft gluon resummation technique at next-to-leading-log accuracy; ~ 3) a direct control over higher-twist effects; and ~ 4) the independence from any specific choice of the $x$-shape of the input parton distributions.

\indent At next-to-leading order we get $\alpha_s(M_Z) = 0.1209 \pm 0.0010 ~ (stat.) \pm 0.0015 ~ (syst.)$ with a significant dependence upon the order of the moment used. Including soft gluon effects we obtain $\alpha_s(M_Z) = 0.1188 \pm 0.0010 ~ (stat.) \pm 0.0014 ~ (syst.)$ with a remarkable better stability against the order of the moment. Our findings compare positively with the total and $DIS$ world averages, $\alpha_s(M_Z) = 0.1183 \pm 0.0027$ and $\alpha_s(M_Z) = 0.119 \pm 0.002 ~ (exp.) \pm 0.003 ~ (th.)$, respectively \cite{update}.

\indent Thus, the new approach we are proposing appears to work quite well and it may represent an interesting tool for providing a precise determination of $\alpha_s(M_Z)$ from $DIS$ data. Directions for future improvements have been discussed. From the experimental side we need an improvement in the quality of the data, at least in the $Q^2$ range from $\approx 5$ to $\approx 20 ~ (GeV/c)^2$, where the largest sensitivity to scaling violations is expected. The upgrade of Jefferson Lab to a 12 GeV electron beam and the use of the $CLAS$ detector might provide such an improvement, offering the possibility to a very precise determination of the {\em experimental} Nachtmann moments of the proton (and deuteron) structure functions for $n > 2$. From the theoretical point of view the comparison of our findings with the recent $NNLO$ results of Refs.~\cite{SY01} and \cite{Alekhin} indicates quite clearly the need of including all the $NNLO$ effects with a consistent resummation of large $n$-logarithms at $NNLL$ accuracy. Moreover, the full singlet and non-singlet evolutions should be considered for the moments with $n > 2$ as well as more Nachtmann moments should be included in our analysis in order to check the stability of the extracted values of $\alpha_s(M_Z)$ against the order $n$ of the moment. Once such a stability is fully reached, the uncertainties on the extracted value of $\alpha_s(M_Z)$ can be further reduced by analyzing the scaling violations of all the moments with $n > 2$ simultaneously.

\section*{Acknowledgments}

\indent The authors gratefully acknowledge G.~Altarelli for a careful reading of the manuscript and many valuable comments.

\end{document}